\documentclass[preprint]{revtex4}
\usepackage{graphicx}
\usepackage{epsfig}
\input{epsf}

\begin{document}

\title{Comparison of metal-amino acid interaction in  Phe-Ag and Tyr-Ag complexes by spectroscopic measurements}
\author{Achintya Singha$^1$, Swagata Dasgupta$^2$ and Anushree Roy$^1$}
\affiliation{$^1$Department of Physics, Indian Institute of
Technology, Kharagpur 721302, India\\
$^2$Department of Chemistry, Indian Institute of Technology,
Kharagpur 721302, India}

\begin{abstract}

In this article, we have compared the metal-amino acid
interactions in Tyr-Ag and Phe-Ag complexes through pH dependent
SERS measurements. By analyzing the variation in relative
intensities of SERS bands with the pH of the amino acid solution,
we have obtained the orientation and conformation of the amino
acid molecules on the Ag surface. The results obtained from our
experimental studies are supported by the the energy minimized
strucutres and the observed charge distributions in different
terminals of the molecules. This, in a way, shows that SERS
measurements not only exhibit the interaction of the amino acid
molecules with Ag clusters but also demonstrate their orientation
around it. We have addressed a long standing query on whether the
amine group is directly attached with Ag surface along with the
carboxylate group and $\pi$-electrons in these systems. In
addition, pH dependent optical absorption and transmission
electron microscopy measurements have been performed to understand
the required conditions for the appearance of the SERS spectra in
the light of the aggregation of metal particles and the number of
hot sites in the sol. Our results confirm that the formation of
hot sites in the sol plays a direct role to form a stable
Ag-ligand complex. Furthermore, the interaction kinetics of
metal-amino acid complexes have been analyzed via both Raman and
absorption
measurements.\\


\end{abstract}

 \maketitle

\section{introduction}

Spectroscopic studies of amino acids adsorbed on  metal surfaces
are important because they can provide an insight into the nature
of metal-biopolymer interactions
\cite{Shoeib:2002,Shu:1986,Bjerneld:2000,Bjerneld:2004,Curley:1988}.
A number of articles have shown that amino acids with small side
chains, for example, Glycine (Gly), Alanine(Ala), interact with
metal colloids through both the amine and carboxylate terminals
\cite{Shu:1986}. On the other hand, for Phenylalanine (Phe),
Tyrosine (Tyr), Tryptophan (Trp), Histidine (His), Leucine (Leu),
Glutamic acid (Glu), and Aspartic acid ( Asp), the adsorption
occurs through the carboxylate group. However, no conclusion is
drawn regarding the amine- silver interaction
\cite{Shu:1986,Bjerneld:2000}.

Surface enhanced Raman scattering (SERS) is an extremely sensitive
technique for monitoring the adsorption of species of very low
concentration and for characterizing the structure and orientation
of the adsorbed species on the rough metal surface
\cite{Kneipp:1999}. Nonresonance SERS is primarily sensitive to
the species attached to the metal surface or present within a few
Angstroms of the metal-dielectric interface. Both electromagnetic
field and adsorbate-surface chemical interactions are responsible
for Raman signal enhancement. The `surface selection rules' and
the SERS intensity of a particular Raman band determine the
average orientation of functional groups of the adsorbates on the
metal surface \cite{Shu:1986}. In conjunction with optical
absorption measurements it has been shown that the single metal
cation does not exhibit any SERS effect. Complexation with an
organic substance is necessary for  signal enhancement
\cite{Teiten:1997}. The colloid activation by chemical treatment
brings the primary metal particles close together building up `hot
sites' that are SERS active in an electromagnetic field of visible
wavelength. `Hot sites' are located at interstitial positions
between metal particles and the number of such positions scale
with the number of particles. In the second step, the activating
molecules of very low concentration are bound to the metal surface
and form the complex \cite{Teiten:1997}. Despite well-documented
limitations, the spectral `richness' in SERS arises not only from
the inherent sensitivity of the process, but also from the ability
of this technique
to determine the interactions in the metal-adsorbate complex and
the relative orientation and conformation of the adsorbed
molecules on the metal surface.


The structural difference between Phe and Tyr is in the presence
of only one -OH group  in the para position of the aromatic ring
in the latter [Fig 1]. In solution, depending on the pH values
[pK$_{1}$ $=$ 2.58 and pK$_{2}$$=$ 9.24] the majority of Phe
molecules assume the structural forms shown as species I-III in
Fig. 1 . Similarly, the structural forms of the majority of Tyr
molecules in solution with different pH values [pK$_{1}$ $=$ 2.20,
pK$_{2}$ $=$ 9.11 and pK$_{3}$ $=$ 10.07 (due to the side chain)]
are shown as species I-IV in Fig. 1. In the sol, Ag particles are
in the ionic Ag$^+$ state because of its low oxidation potential.
The electrostatic interaction between the above species of amino
acid molecules with positively charged Ag particles in solution,
is responsible for the SERS effect. However, the structure of the
molecules also plays an important role in determining the
intensities of different SERS bands. Thus, a pH dependent study of
SERS spectra with a comparison between the results for Phe and Tyr
is expected to determine the metal-amino acid interaction and
orientation of the adsorbed species on the metal surface.

The SERS spectra of Phe and Tyr adsorbed on Silver (Ag) colloids
are available in the literature \cite{Bjerneld:2000,Rava:1984}. We
find that though the structure of these molecules are very similar
(as mentioned above), their SERS spectra are widely different. To
understand the interaction of these amino acids with Ag colloids
we have studied the pH dependent optical absorption and SERS
spectra of both Phe-Ag and Tyr-Ag complexes. From the relative
change in intensities of the vibrational bands in SERS spectra we
have compared the interaction, conformation and the orientation of
these molecules on Ag particles.  Section II covers the sample
preparation techniques, which we have followed and other details
regarding the instruments, which we have used for various
measurements. In Section III, we have compared the possible
surface geometry of the adsorbed Phe and Tyr on the surface of the
Ag  particles via pH dependent SERS measurements. Our conjecture
from these experimental observations has been further supported by
the surface geometry expected from the electrostatic interaction
between Ag$^+$ and Phe or Tyr, obtained from the estimated atomic
charge distribution at the different terminals of the amino acid
molecules. In this section we have also shown the role of
formation of hot sites to exhibit SERS spectra for metal-amino
acid complex through simultaneous pH dependent optical absorption
and SERS measurements. The interaction kinetics between the
adsorbate and Ag colloids has been discussed in Section IV.
Finally, in Section V we have summarized our results with a few
concluding remarks.

\section{Materials and Methods}
Silver nitrate (AgNO$_{3}$), sodium borohydride (NaBH$_{4}$),
sodium hydroxide (NaOH), and hydrochloric acid (HCl) of analytical
reagent grade (SRL, India), were used to prepare the Ag sol. Amino
acids were also obtained from the above company. A colloidal
silver solution was prepared in deionized water according to the
method described by Creighton et al. \cite{Creighton:1979}. This
method essentially uses the reduction of 10$^{-3}$ M AgNO$_{3}$ by
an excess amount of 10$^{-3}$ M NaBH$_{4}$.  10$^{-3}$ M
AgNO$_{3}$ was added dropwise to 10$^{-3}$ M NaBH$_{4}$ (placed in
an ice bath under quick stirring condition) maintaining a 1:3
volume ratio. Stirring for 20 minutes was necessary to stabilize
the colloidal solution. Later, it was left at room temperature for
approximately 1 hour. The excess NaBH$_{4}$ evaporated and the
remaining solution became transparent yellow in color. 10$^{-3}$ M
solution of Phe and Tyr were prepared in deionized water. The pH
of the solution was adjusted by using 1 M HCl and 1 M NaOH. For
SERS measurements Phe and Tyr solutions of different pH were added
to the Ag sol. For all SERS experiments the volume ratio of Ag sol
to amino acid solution was maintained at 9:1.

The SERS spectra were measured using a 488 nm Argon ion laser as
an excitation source. The spectrometer is equipped with 1200
g$/$mm holographic grating, a holographic super-notch filter, and
a Peltier cooled CCD detector. UV-Visible spectra were measured by
Spectrascan UV 2600 (Chemito). All absorption spectra are recorded
2 minutes after addition of the amino acid to the  sol. Samples
for Transmission Electron Microscopy  were deposited onto 300 mesh
copper TEM grids coated with 50 nm carbon films. The excess water
was allowed to evaporate in air. The grids were examined in JEOL
2010 microscope with Ultra-High Resolution (UHR)  microscope using
a LaB$_6$ filament operated at 200 kV.

The amino acid zwitterionic structures were drawn and energy
minimized using SYBYL 6.92 (Tripos Inc., St. Louis USA). The
structures were rendered in Pymol (Delano Scientific LLR, USA).
The atomic charge distribution of the molecules has been estimated
using the Gasteiger-H$\ddot{u}$ckel method.

\section{Adsorption of  Phenylalanine (Phe) and Tyrosine (Tyr) on
Silver colloidal surface: Optical absorption and SERS
measurements}

\subsection{Phenylalanine (Phe) : Results and Discussion}


The optical absorption spectra (OAS) of Phe at different pH in a
colloidal solution of Ag are shown in Fig. 2. The Ag sol absorbs
light at $\lambda_{max}$ = 392 nm (A) [Fig. 2]. This band is
associated with the dipolar surface plasmon of small, single,
spherical particles of Ag. Transmission electron micrograph (TEM)
of Ag sol shows formation of well-dispersed  spherical Ag
particles [Fig. 3(a)]. The average diameter of the Ag particles
estimated from the $\lambda_{max}$ in OAS in Fig. 2 and TEM image
in Fig. 3(a) is $\sim$ 90 \AA.

Addition of Phe to the sol results in a decrease in intensity of
peak A (due to a single silver particle), along with an appearance
of a new band at around 561 nm (B) due to aggregated Ag particles.
In Fig. 2 we have shown the change in the optical absorption
spectrum of Ag sol with addition of Phe of pH varying from 1.5 to
12.5. The change in the ratio of intensity of peak B to that of
peak A with variation in pH is shown in Fig. 4. The nature of
aggregation of the silver particles, with addition of Phe of
different pH, as obtained from TEM images, is shown in Fig. 3(b)
to Fig. 3(d). Addition of  Phe at a very low pH (below pH 2) to
the Ag sol results in an immediate agglomeration and precipitation
of Ag particles (not shown in the figure). One may also note the
green curve in Fig. 2, where any signature of colloidal Ag
particle is nearly absent. Between pH 2 and pH 3.5 the metal
particles agglomerate appreciably. However, they are not big
enough to precipitate immediately [Fig. 3(b) for pH 2]. For Phe
with medium pH in Ag sol, one observes clustering of Ag particles
in solution without much agglomeration or precipitation [Fig. 3(c)
for pH 8 of Phe]. Addition of Phe at very high pH values (above 9)
prohibits the clustering of Ag particles, which now spread out in
solution [Fig. 3(d)].




\begin{table}[htbp]
\caption{Assignment of vibrational bands for Phe-C (Raman
spectrum) and Phe-Ag complex(SERS spectra) }
\begin{tabular}{|c|c|c|} \hline
Assignment  &  Phe - C (cm $^{-1}$)  & SERS  Phe (cm $^{-1}$) \\
\hline
Benzene ring breathing mode ($\nu_1$) & 1008  &  1000 \\
COO$^-$ rocking, wagging, bending & 500-700 & 500-700 \\
$\nu_{C-C}$ of Benzene ring & 901 &  - \\
C-H vibration in benzene ring & 3080 & 3065 \\
C-COO$^-$      & 929 & 936 \\
CN             & 1036 &  1035 \\
NH$_{3}^{+}$   & 1595 & 1582 \\
CH             & 1251 & 1235 \\
CH             & 2976 & 2915 \\
COO$^{-}$      & 1413 & 1375 \\
CH$_{2}$ wagging & 1332 &    - \\
CH$_{2}$ deformation & 1400 &  - \\ \hline
\end{tabular}
\end{table}

In  Fig. 5 we have shown the Raman spectrum of an aqueous Phe
solution of 0.5 M ( mentioned as Phe-C in the rest of the article)
at pH 6.5 within the spectral window of 500- 3200 cm$^{-1}$ (after
subtracting the broad background of water in the range 1500-1800
cm$^{-1}$ and 2800-3200 cm$^{-1}$). The assignments of bands in
Phe-C, summarized in Table I, are based on the available Raman
data for Phe \cite{Stephen:1989,Rava:1985}. The most intense band
at 1008 cm$^{-1}$ has been assigned to the benzene ring breathing
mode ($\nu_{1}$). Other less intense but prominent bands appear at
901 cm$^{-1}$, 929 cm$^{-1}$, 1036 cm$^{-1}$, and 1595 cm$^{-1}$
due to $\nu_{C-C}$, C-COO-, CN and NH$_{3}^{+}$ vibrations,
respectively. The bands belonging to $\nu_{CH}$, $\nu_{C00^{-}}$,
CH$_2$ wagging, CH$_2$ deformation are weak and observed at 2976
(also at 1251), 1413, 1332  and 1400 cm$^{-1}$, respectively.
COO$^{-}$ rocking, wagging, and bending modes appear between 500
and 700 cm$^{-1}$. The feature at 3080 is due to the breathing
motion of the hydrogen atoms in the benzene ring (C-H vibration).

SERS spectra (solid lines) of 1$\times$ 10$^{-3}$ M Phe, 2-3
orders of magnitude more dilute than the concentration typically
required for nonoresonance Raman scattering, with pH from 1.5 to
12.5 are shown in Fig. 6. We have recorded  spectra of Phe-Ag
complexes by changing the pH values by 1.0 unit over this range.
However, in Fig. 6, we have shown a few characteristic spectra.
SERS peak positions are also listed in Table 1. The Raman spectrum
of aqueous Phe solution with the low concentration (0.001 M) at pH
7 is shown by the dotted line in the same figure.


To understand the interaction of the adsorbed species on the
surface of Ag particles, we will make use of both SERS and
absorption spectra. It is clear from Fig. 5 and Fig. 6 that
intensities of the above mentioned vibrational modes change in
SERS spectra with variation in pH. It is also interesting to note
that below the pH value of 4.5 and above the pH value of 10 the
SERS spectrum is extremely weak for the complex. As we mentioned
before, upon addition of Phe of very low pH (below 3.5, value of
pK$_1$ = 2.58) the Ag particles agglomerate and then precipitate
or form bigger particles [see Fig. 2 green curve and also Fig. 3
(b)]. The high pH value (above 9.5, pK$_2$ = 9.24) results in a
spreading of Ag particles, as shown in Fig. 3(d). The slight red
shifted broad navy blue spectrum  for pH 12 in Fig. 2 indicates
the absence of agglomerated Ag particles alone in the sol.  Under
both these conditions the number of hot sites decreases in
solution, which in turn results in a decrease in the intensity of
the SERS spectrum. In other words, the inefficient SERS is related
to the failure of colloid activation due to insufficient
complexation of Phe to the Ag surface. Intensities of the
vibrational bands in SERS spectra are quite prominent, though of
varying intensities, for the pH values ranging from 4.5 to 9.5.
Thus, from Fig. 3 and Fig. 4, we conclude that for efficient SERS
the aggregation of Ag colloidal particles (Intensity ratio of B to
A in optical absorption spectra) plays an important role via the
formation of hot sites.

The SERS spectra at pH below 3.5, between 3.5 and 9.5, and above
9.5 should correspond to species I, II and III respectively, shown
in Fig. 1(a). For Benzene molecule \cite{Shu:1986} SERS surface
selection rule states that vibrational bands which draw the
intensity of Raman polarizability component $\alpha_{zz}$ is most
intense, where $z$ is the surface normal. If a benzene molecule is
absorbed, such that the ring lies parallel  to the metal surface,
$\alpha_{zz}$ contributes only to its two A$_{1g}$ modes $\nu_{1}$
(at 992 cm$^{-1}$) and $\nu_{2}$ (at 3056 cm$^{-1}$). The maximum
contribution to the $\alpha_{zz}$ component of the polarizability
of benzene comes from the $\pi$- electrons. Thus, $\nu_1$, which
perturbs the carbon ring directly, plays a more prominent role in
affecting $\alpha_{zz}$  than $\nu_{2}$, which involves only
hydrogen atoms \cite{Shu:1986}. However, if the benzene ring is
perpendicular to the metal surface rather than lying parallel, one
expects $\nu_2$ would be more intense than $\nu_{1}$. In aqueous
Phe these two peaks of the aromatic ring structure appear at 1008
and 3080 cm$^{-1}$. In SERS spectra these modes are obtained at
1000 and 3065 cm$^{-1}$. Above the pH value 3.5, the intensity of
the band at 1000 cm$^{-1}$ increases with increase in pH of Phe
till the pH value reaches 9.5. Beyond this value of pH  the 1000
cm$^{-1}$ band disappears [Fig. 7a]. The low intensity ratio of
the band at 1000 cm$^{-1}$ and  3056 cm$^{-1}$ compared to what we
observe for Phe-C (Fig. 5) indicates that the aromatic ring of Phe
is absorbed \emph{nearly} parallel to the surface of the Ag
colloids. With increase in pH there is a conformational change in
the molecule; the attached molecule tilts more and more towards
the surface (flatter) as indicated by the increase in ratio of
$I(\nu_{1})/I(\nu_{2})$, shown in  Fig. 7(b). The above intensity
variation and shift in the frequency of these bands in the SERS
spectrum with respect to their frequencies in the spectrum of the
solution, indicate that the $\pi$-system of Phe participates in
complex formation with Ag.



In Fig. 6, a relatively strong band at 1375 cm$^{-1}$ beyond pH
value 3.5 (note that pK$_1$ =2.58 for Phe) and the variation in
intensity of this band with pH indicates that the carboxylic
group, -COO$^{-}$, upon deprotonation is clearly absorbed on the
surface of the Ag colloids [Fig. 8(a)]. There are several weak
bands between 500 and 750 cm$^{-1}$ due to different COO$^{-}$
vibrational modes.  In SERS spectra of Phe-Ag the CH- stretching
vibration appears at 2915 cm$^{-1}$. The $\nu$CH$_2$ vibrational
modes are absent in SERS spectra, indicating that the methylene
group is relatively far from the Ag surface.  Though, it is not
expected that the NH$_{3}^{+}$ group be directly attached to
Ag$^+$ colloids in the sol, the relatively strong band due to
asymmetrical deformation of NH$_{3}^{+}$ at 1582 cm$^{-1}$
indicates that amine group of the molecule stays relatively close
to the Ag surface. The intensity of this band drops down with
increase in pH as NH$_{3}^{+}$ transforms to NH$_{2}$ [Fig. 8(b)].
The relatively strong band due to $\nu$CN at 1036 cm$^{-1}$, and
its increase in intensity with increase in pH [Fig. 8(c)] can be
explained by assuming the proximity of  amine group to the silver
surface.

\subsection{Tyrosine (Tyr) : Results and Discussion}


The optical absorption spectra of Tyr at different pH in colloidal
solution of Ag are shown in Fig. 9. As observed in case of Phe
[Fig. 2], the addition of Tyr to the sol results in the appearance
of a new band at around 520 nm (B1) due to aggregation of Ag sol
along with a decrease in intensity of peak A (due to a single
particle). The variation in the ratio of the intensity of peak B1
to that of peak A with change in pH is shown in Fig. 10. The
nature of aggregation of the silver particles with addition of Tyr
of different pH, as obtained from TEM images are shown in Fig. 11
(a) - Fig. 11(c).



In Fig. 12 we have shown the Raman spectrum of an aqueous Tyr
solution of 0.5 M (Tyr-C) at pH 6.0 (after subtracting the
luminescence background). The assignment of vibrational bands in
Tyr-C has been summarized in Table II. The most intense band at
828 cm$^{-1}$ has been assigned to the breathing mode of the
aromatic ring structure of Tyr \cite{Rava:1984}. Other less
intense but prominent bands, appear at 931 cm$^{-1}$, 1304
cm$^{-1}$ and 2967 cm$^{-1}$, are due to C-COO$^-$, CH$_2$ wagging
and CH vibrations, respectively. The bands belonging to $\nu_{CH}$
vibrations in the ring appears at 3061 cm$^{-1}$.



\begin{table}[htbp]
\caption{Assignment of vibrational bands for Tyr-C (Raman
spectrum)) }
\begin{tabular}{|c|c|} \hline
Assignment  &  Tyr - C (cm $^{-1}$) \\ \hline
Ring breathing mode  & 828 \\
C-COO$^-$            & 931  \\
CH$_{2}$ wagging     & 1304  \\
CH                   & 2967  \\
CH vib.      in ring & 3061    \\ \hline
\end{tabular}
\end{table}


The SERS spectra of Tyr (1$\times$ 10$^{-3}$ M) with pH 1.5 to
12.5 are shown in Fig. 13. Raman spectrum of aqueous Tyr solution
of same concentration at pH 7 is shown by a dotted line in the
same figure. It is interesting to note that though the molecular
structure of Tyr is similar to that of Phe except for an -OH group
in the para position of the aromatic ring, the SERS spectrum is
entirely different from that of Phe. In aqueous solution, Tyr
becomes Tyrosinate, TyrO$^-$, at pH of 10.07 (pK$_3$). However,
Tyr deprotonates upon adsorption, ie. they are absorbed as
TyrO$^-$ on metal surface (It has been shown that SERS spectra of
Tyr-Ag should cover much narrower spectral region)
\cite{Bjerneld:2000,Rava:1984}. We assign the broad and unresolved
band between 1200 and 1500 cm$^{-1}$ in Fig. 13, measured with
long integration time (5 minutes as in our case) for the range of
pH value from 3.5 to 9.5, to temporal averaging (averaging of
strongly fluctuating contribution to overall intensity profile) of
Tyr SERS spectrum. The reason is the following : In addition to
the direct coordination of the aromatic ring with Ag$^+$, TyrO$^-$
can be adsorbed on the Ag surface either via the carboxylic acid
group or via both the amino and carboxylic group or via the phenol
hydroxyl group. The ensemble average of all these complexations
results in a broad band. This has been also been shown from
quantum chemical estimates of vibrational energies using
Hartree-Fock self-consistent field molecular orbital calculation
for TyrO$^-$-Ag complex \cite{Bjerneld:2000}.
In Fig. 14 we have
shown the variation in the intensity of the peak at  1390
cm$^{-1}$ with the pH of the Tyr solution. In the acidic region (
pH$<$4.5) an extensive aggregation [Fig. 11(a)] of the colloid
silver leads to a complete loss of the SERS signal.The decrease in
SERS spectral intensity for higher pH at 10.5 and above is due to
spreading out of the Ag particles in sol, which decreases the
number of hot sites necessary for SERS [Fig 11 (c)]. Following the
discussion on Phe-Ag complex, the role of agglomeration of Ag
particles (formation of the number of hot sites in metal sol) to
exhibit efficient SERS of Tyr-Ag complex is also clear from the
variation in intensity ratio of peak A and peak B1 with pH in
absorption spectra as shown in Fig. 10 and also from the TEM
images shown in Fig. 11.


\subsection{Comparison of interaction in Phe-Ag and Tyr-Ag
complex}

From the above experimental results and the corresponding
discussion we observe that the Phe and Tyr molecules behave very
differently while forming complexes with Ag molecules. For Phe the
aromatic ring lies nearly flat on Ag surface. Both carboxylic and
amine groups are important in adsorption of Phe on metal
particles; whereas, the methylene (CH$_2$) group is relatively
inert. This signifies that the aromatic ring, carboxylic and amine
group of Phe lie in close proximity of metal particles, whereas
the methylene group stays away. On the other hand, Tyr adsorbs on
Ag surface as Tyrosinate. At any instant of time, interaction of
metal ion either with carboxylic group or with amine group or with
oxygen in the side chain can be more probable than the other two
interactions.

To support our above conjecture, we have looked into the atomic
charge distribution at the different terminals of Phe and Tyr
molecules in the energy minimized zwitterionic forms (which bind
with Ag ion to form a complex), shown in Fig. 15 (a) and 15(b).
From the hybrid density functional theory, it has been shown that
while forming a complex with Phe or Tyr, the Ag$^+$ preferentially
occupies the cavity formed in the molecular structure, shown in
Fig. 15 \cite{Shoeib:2002a}. If one takes into account only the
electrostatic affinity between Ag and Phe/Tyr molecule, from the
atomic charge distribution, shown in Fig. 15, it appears that the
following interactions are possible for both Ag-Phe or Ag-Tyr
complexes  - (a)it is unlikely for the NH$_{3}^{+}$ group to be
directly attached with the positively charged Ag surface, though
it is in close proximity of the Ag$^+$,  (b) the CH$_2$ group is
away from Ag$^+$, which results in a weak electrostatic
interaction between them, (c) the negative charge distribution on
oxygen atoms of the carboxylic acid group and ring structure
indicate  the possibility of a strong electrostatic interaction
with the positive metal ion (d) the aromatic ring is parallel to
the surface of the Ag$^+$. From a more careful analysis we see
that in Phe the charges on the oxygen atoms of the carboxylic
group are - 0.56 units, whereas, those on Tyr are - 0.29 and -
0.36 units. The side chain oxygen atom of Tyr has a charge of -
0.34 units. We believe the relatively large negative charge on
oxygen terminals in Phe holds the positive metal ion firm in its
position via Coulombic interactions. However, the relatively less
negative charge at the carboxylic oxygen terminals and at the side
chain result in a smeared position of the Ag$^+$ in the Tyr cavity
[Fig. 15 (b)] compared to what we observe in case of the Phe-Ag
complex [Fig. 15 (a)]. This explains the weak and broad SERS band
for Tyr-Ag complex. It is also interesting to note that except in
these terminals, the charge distributions are nearly the same on
all other atoms in Phe and Tyr.

\section{Kinetic measurements on  Ag-Phe and Ag-Tyr interaction}
We have seen in the previous section that the formation of hot
sites due to the optimum agglomeration of Ag colloidal particles
plays a direct role to form stable Ag-ligand complex. In this
section, we have confirmed the same fact by measuring the
interaction kinetics of Phe-Ag and Tyr-Ag complexes by SERS and
absorption measurements. To study the kinetics of the interaction
we have recorded the intensity of optical absorption peak  at 535
nm (peak B due to agglomerated Ag in Fig. 2) for Phe-Ag complex
and 490 nm (peak B1 in Fig. 9) for Tyr-Ag complex with time. The
variations in intensities of these two peaks with time are shown
in Fig. 16 (a) and 16 (b), respectively. We clearly observe two
different decay rates for both the peaks. For the first 120 sec
the decay process is very fast compared to what is observed later.
The decay time constants are 58 ($\tau_1$) sec and 178 ($\tau_2$)
sec for Phe and 63 ($\tau_1$) sec and 100 ($\tau_2$) sec for Tyr.


To correlate the optical absorption and SERS measurements, we
studied the kinetics of SERS spectra of 5 $\times$ 10$^{-3}$ M
ligand- Ag mixture at pH 8 (about which we get maximum SERS
intensity for the most intense peaks). Each spectrum was taken for
2 minutes with 1 minute interval. The exponential drop in
intensities of the strongest band at 1000 cm$^{-1}$ for Phe and
the band at 1392 cm$^{-1}$ in Tyr are shown in Fig. 17(a) and
17(b). The decay constants are estimated to be 213 sec and 113
sec, respectively. These values of the time constants match
reasonably well with the values of $\tau_{2}$, which we have
obtained from optical absorption measurements. Thus, we confirm
that the metal-amino acid interaction is directly related to the
aggregation of the Ag colloids. Here we would like to mention that
because of experimental limitations the absorption spectra in Fig.
2 and Fig. 10 or SERS spectra in Fig. 6 and Fig. 14 are recorded 2
minutes after addition of amino acids in Ag sol. Thus, the effect
of initial kinetics of interaction, at time scale $\tau_1$, is not
observed in these spectra.



However, a careful observation of the initial change in colour of
the sol with addition of amino acid indicates fast kinetics of the
interaction between ligand molecules and Ag sol. The as-prepared
Ag sol is yellow in colour and is stable for more than 24 hrs at 4
$^\circ$C. The agglomeration of the particle after addition of
amino acids changes the colour from yellow to pink to blood red as
shown in  Fig. 18 (a) for Phe and yellow to amber to olive green
as shown in Fig. 18 (b) for Tyr. It is to be noted that the change
in colouration of the solution is rapid (distinct by naked eye)
for first $\sim$ 60  seconds for both Phe-Ag and Tyr-Ag. This time
scale matches with value of $\tau_1$ obtained from kinetic studies
by absorption measurements. This initial change in color may be
due to the increase in the size of the particles from single
particle to  doublets, triplets and higher multiplets. Gradually,
with time, the colour of the solution fades out indicating the
breakdown of agglomerated particles into smaller fragments.

\section{Summary}

Though structurally close, Phe and Tyr, behave very differently,
when they form complexes with Ag colloids. To understand this
difference, we have compared the metal-amino acid interaction in
Phe-Ag and Tyr-Ag complexes by pH dependent SERS measurements.
From the intensity variations of the vibrational bands in SERS
spectra with the pH of the adsorbates, we have proposed the
relative orientation and interaction of the adsorbed molecules on
the Ag surface. We have addressed a long standing query, as to
whether the amine group is directly attached with Ag surface along
with carboxylate group in these systems.  Using
Gasteiger-H$\ddot{u}$ckel method to estimate the atomic charge
distribution at different terminals of the zwitterionic amino acid
molecules, we have noted that, though the structures of Phe and
Tyr are quite similar, (except an -OH group at the para position
of the ring in the latter), the atomic charge distributions at the
oxygen sites of these molecules, are quite different. If we assume
only the electrostatic interaction between different terminals of
the adsorbates with Ag$^+$, the above charge distribution confirms
the interaction and orientation of the amino acid molecules in the
complexes, which we have proposed from SERS measurements.

In addition, the appearance of SERS bands have been explained with
the help of pH dependent TEM and optical absorption measurements.
We have seen that formation of hot sites via optimum aggregation
of Ag particles induced by Phe/Tyr results in an increase in
intensity of the optical absorption band of the Ag aggregates and,
concurrently, an increase in SERS band intensities. We have also
shown that the state of aggregation is a key parameter in SERS. We
have  discussed the reaction kinetics of this interaction process
using spectroscopic measurements.

\section{Acknowledgements}
 Authors thank P.V. Satyam and J. Ghatak, IOP,
Bhubaneshwar, India, for their assistance in the TEM work. AR
thanks Department of Science and Technology, India, and Third
World Academy of Science, ICTP, Italy for financial assistance.

\pagebreak

\noindent
\textbf{ Figure Captions}

\noindent Figure 1. Different ionic species of Phe and Tyr at
different pH values

\noindent Figure 2. Absorption spectra of Phe -Ag mixture at
different pH values.

 \noindent
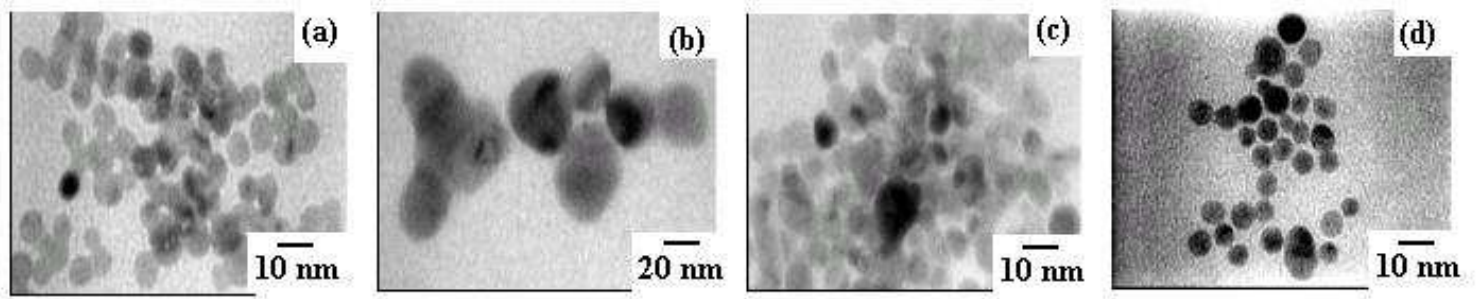
 Figure 3. Transmission electron micrograph of (a) Ag sol and Phe-Ag mixture at (b) pH 2 (c) pH 8, and (d) pH
 11.

\noindent
Figure 4. Variation of intensity ratio of peak B to peak
A in optical absorption spectra for Phe-Ag complex.

\noindent
Figure 5. Raman spectrum of Phe in solution at
concentration 0.5M.

\noindent
Figure 6. Raman spectrum (dotted) of Phe and SERS
spectra (solid lines) of Phe-Ag complex (1$\times$ 10$^{-3}$ M) at
different pH of Phe.

\noindent Figure 7. Intensity variation of SERS bands at (a) 1000
cm$^{-1}$, (b) intensity ratio I($\nu_1$)/I($\nu_2$) with pH
variation from 1.5 to 12.5 for Phe-Ag complex.

\noindent Figure 8. Intensity variation of SERS bands at (a) 1375
cm$^{-1}$, (b)1582 cm$^{-1}$ and (c) 1036 cm$^{-1}$ with pH
variation from 1.5 to 12.5 for Phe-Ag complex.

\noindent
Figure 9. Absorption spectra of Tyr -Ag mixture at
different pH.

\noindent
Figure 10. Variation of intensity ratio of peak A and
peak B1 in optical absorption spectra for Tyr-Ag complex.

\noindent Figure 11. Transmission electron micrograph of  Tyr-Ag
mixture at (a)pH 2, (b) pH 7.5, and (c) pH 11.

\noindent Figure 12. Raman spectrum of Tyr in solution at
concentration 0.5M.

\noindent Figure 13. Raman spectrum of Tyr (dotted) and SERS
spectra (solid lines) of Tyr-Ag complex (1$\times$ 10$^{-3}$ M) at
different pH.

 \noindent Figure 14.
Intensity variation of Raman bands at 1390 cm$^{-1}$ in Tyr with
change in pH  from 1.5 to 12.5.

\noindent Figure 15. Atomic charge distribution on (a) Phe and (b)
Tyrosinate, obtained by Gasteiger-H$\ddot{u}$ckel method.

\noindent Figure 16. Kinetic study of (a) the feature at 535 nm in
optical absorption spectra of Phe -Ag mixture and (b) the feature
at 490 for Tyr-Ag complex. Insets of the figures show same for
first 2 min. duration

 \noindent
 Figure 17.
Kinetic study of the most intense peak in SERS spectra for
(a)Phe-Ag complex and (b) Tyr-Ag complex.

\noindent Figure 18. Change in colour of (a) Phe-Ag complex and
(b) Tyr-Ag complex with time

\end{document}